\documentclass{emulateapj}

\usepackage{amssymb}
\usepackage{amsmath}
\usepackage{graphicx}
\usepackage{natbib}
\usepackage{txfonts}

\bibpunct{(}{)}{;}{a}{}{,} % to follow the A&A style

\def\roma{1}
\def\ita{2}
\def\icra{3}
\def\uff{4}

\begin{document}

\title{Dynamical instability of white dwarfs and breaking of spherical symmetry
under the presence of extreme magnetic fields}
\author{J. G. Coelho\altaffilmark{\roma,\ita,\icra},
                                R. M. Marinho\altaffilmark{\ita},
                                M. Malheiro\altaffilmark{\ita},
                                R. Negreiros\altaffilmark{\uff},
                                D. L. C\'aceres\altaffilmark{\roma,\icra},
				J. A. Rueda\altaffilmark{\roma,\icra},
				and
        R. Ruffini\altaffilmark{\roma,\icra}
        }

\altaffiltext{\roma}{Dipartimento di Fisica and ICRA, 
                     Sapienza Universit\`a di Roma, 
                     P.le Aldo Moro 5, 
                     I--00185 Rome, 
                     Italy}
                                          
\altaffiltext{\ita}{Departamento de F\'isica, Instituto Tecnol\'ogico de Aeron\'autica, ITA,
S\~ao Jos\'e dos Campos, 12228-900, SP, Brazil}    

\altaffiltext{\icra}{ICRANet, 
                     P.zza della Repubblica 10, 
                     I--65122 Pescara, 
                     Italy}
\altaffiltext{\uff}{Instituto de F\'isica, Universidade Federal Fluminense, UFF,
Niter\'{o}i, 24210--346, RJ, Brazil}

\altaffiltext{}{jaziel.coelho@icranet.org,m.malheiro@ita.br,jorge.rueda@icra.it}

\begin{abstract}
Massive, highly magnetized white dwarfs with fields up to $10^9$~G have been observed and theoretically 
used for the description of a variety of astrophysical phenomena. Ultramagnetized white dwarfs with uniform
interior fields up to $10^{18}$~G, have been recently purported to obey a new maximum mass
limit, $M_{\rm max}\approx 2.58~M_\odot$, which largely overcomes the traditional Chandrasekhar 
value, $M_{\rm Ch}\approx 1.44~M_\odot$. Such a much larger limit would make these astrophysical
objects viable candidates for the explanation of the superluminous population of type Ia supernovae.
We show that several macro and micro physical aspects such as gravitational, dynamical stability, 
breaking of spherical symmetry, general relativity, inverse $\beta$-decay, and pycnonuclear fusion 
reactions are of most relevance for the self-consistent description of the structure and assessment
of stability of these objects. It is shown in this work that the first family of magnetized white dwarfs 
indeed satisfy all the criteria of stability, while the ultramagnetized white dwarfs are very unlikely to
exist in nature since they violate minimal requests of stability. Therefore, the canonical Chandrasekhar 
mass limit of white dwarfs has to be still applied.
\end{abstract}


\maketitle

%%%%%%%%%%%%%%%%%%%%%%%%%%%%%%%%%%%%%%%%%%%%%%%%%%%%%%%%%%%%%%%%%%%%%%%%%%%%%%%%%%%%%%%%%%%%%%%%%%
%%%%%%%%%%%%%%%%%%%%%%%%%%%%%%%%%%%%%%%%%%%%%%%%%%%%%%%%%%%%%%%%%%%%%%%%%%%%%%%%%%%%%%%%%%%%%%%%%%
\section{Introduction}\label{sec:1}
%%%%%%%%%%%%%%%%%%%%%%%%%%%%%%%%%%%%%%%%%%%%%%%%%%%%%%%%%%%%%%%%%%%%%%%%%%%%%%%%%%%%%%%%%%%%%%%%%%
%%%%%%%%%%%%%%%%%%%%%%%%%%%%%%%%%%%%%%%%%%%%%%%%%%%%%%%%%%%%%%%%%%%%%%%%%%%%%%%%%%%%%%%%%%%%%%%%%%

There is an increasing interest of the astrophysics community on highly magnetized white dwarfs (HMWDs) both from the theoretical and observational point of view.
HMWDs with surface fields from $10^6$~G up to $10^{9}$~G have been confirmed observationally mainly via Zeeman splitting of the spectral 
lines \citep{2009A&A...506.1341K,2010yCat..35061341K,2010AIPC.1273...19K,2013MNRAS.429.2934K}. Besides
their high magnetic fields, most of them have been shown to be massive, and responsible for the high-mass peak at $1~M_\odot$ 
of the white dwarf distribution; for instance: REJ 0317--853 with $M \approx 1.35~M_\odot$ and
$B\approx (1.7$--$6.6)\times 10^8$~G \citep{1995MNRAS.277..971B,2010A&A...524A..36K}; PG 1658+441
with $M \approx 1.31~M_\odot$ and $B\approx 2.3\times 10^6$~G \citep{1983ApJ...264..262L,1992ApJ...394..603S};
and PG 1031+234 with the highest magnetic field $B\approx 10^9$~G \citep{1986ApJ...309..218S,2009A&A...506.1341K}.

From the theoretical point of view, massive, rapidly rotating, HMWDs have been proposed as an alternative scenario to the magnetar model
for the description of soft-gamma repeaters (SGRs) and anomalous X-ray 
pulsars (AXPs) \citep{1988ApJ...333..777M,paczynski90,2012PASJ...64...56M,2013A&A...555A.151B,2013ApJ...772L..24R,Coelho2013aarxiv,Coelho2013barxiv,Coelho2014}.
Such white dwarfs were assumed to have fiducial parameters $M = 1.4~M_\odot$, $R = 10^8$~cm, $I = 10^{49}$~g~cm$^2$, and 
magnetic fields $B\sim 10^7$--$10^{10}$~G, inferred using a typical oblique rotating magnetic dipole model and the
observed rotation periods, $P\sim (2$--$12)$~s, and spindown rates, $\dot{P}\sim 10^{-11}$~s/s.

Super-Chandrasekhar white dwarfs with high magnetic fields have been recently used to explain some properties of supernovae. 
Their masses overcome the traditional Chandrasekhar limit, 
\begin{equation}\label{eq:chandralimit}
M_{\rm Ch}=2.015\frac{\sqrt{3\pi}}{2}\frac{m^3_{\rm Pl}}{(\mu_e m_H)^2}\approx 1.44~M_\odot,
\end{equation}
where $\mu_e\approx 2$ is the mean molecular weight per electron, $m_H$ the mass of hydrogen atom, and $m_{\rm Pl}=\sqrt{\hbar c/G}$ is the Planck mass. 

Since such objects should be metastable, the magnetic braking of magnetized, $B\sim 10^6$--$10^8$~G, Super-Chandrasekhar white dwarfs with $M\sim 1.5~M_\odot$, 
have been adopted to explain the delayed time distribution of type Ia supernovae \citep[see][for details]{2012MNRAS.419.1695I}: 
the explosion would be delayed for a time typical of the spindown time scale due to magnetic braking, providing the result
of the merging process is a magnetized Super-Chandrasekhar white dwarf rather than a sub-Chandrasekhar one.

Super-Chandrasekhar white dwarfs have been also claimed to be able to explain the observed properties of some peculiar superluminous Ia supernovae,
which need white dwarf progenitors with masses $(2.1$--$2.8)~M_\odot$, depending on the amount of nickel needed to 
successfully explain both the low kinetic energies and high luminosity of these
supernovae \citep{2006Natur.443..308H,2007ApJ...669L..17H,2009ApJ...707L.118Y,2010ApJ...713.1073S,2011MNRAS.410..585S,2011MNRAS.412.2735T}. 

Following this idea, \citet{2013PhRvL.110g1102D} recently purported that the effects
of a quantizing strong and uniform magnetic field on the equation of state (EOS) of a white dwarf, would
increase its critical mass up to a new value $M_{\rm max}\approx 2.58~M_\odot$, significantly exceeding 
the Chandrasekhar limit (\ref{eq:chandralimit}). This result would imply these objects as viable progenitors
of the above superluminous type Ia supernovae. This new mass limit would be reached, in principle, for extremely 
large interior magnetic fields of the order of $10^{18}$~G.

Therefore, since HMWDs are acquiring a most relevant role in astrophysical systems, it is of major importance
to assess the validity of the assumption of the existence in nature of these objects on theoretical grounds. 
The effect of chemical composition, general relativity, and inverse $\beta$-decay on the determination of the 
maximum stable mass of non-rotating white dwarfs was studied both qualitatively and quantitatively in \citep{2011PhRvD..84h4007R}.
The extension to the uniformly rotating case including the analysis of rotational instabilities (mass-shedding and secular 
instability), inverse $\beta$-decay, and pycnonuclear reactions was analyzed in \citep{2013ApJ...762..117B}. It was shown in 
the latter that white dwarfs might have rotation periods as short as $0.3$~s. However, the above theoretical analyses considered the white dwarf as unmagnetized. 

We show in this work that several macro and micro physical aspects such as gravitational, dynamical stability, breaking of spherical symmetry, general
relativity, inverse $\beta$-decay, and pycnonuclear fusion reactions are relevant for the self-consistent description 
of the structure and assessment of stability of ultramagnetized white dwarfs. Our analysis leads to two major conclusions: 

(1) in the particular case of sub-Chandrasekhar white dwarfs (or slightly exceeding the Chandrasekhar limiting value e.g. by rotation)
with surface magnetic fields in the observed range, i.e.~$B\sim 10^6$--$10^{9}$~G, the unmagnetized approximation for the description
of the structure parameters, e.g.~mass and radius, is approximately correct and therefore, the results
of \citet{2011PhRvD..84h4007R,2013ApJ...762..117B}, can be safely used for the static and uniformly rotating cases, respectively;

(2) the new mass limit $M_{\rm max}\approx 2.58~M_\odot$ \citep{2013PhRvL.110g1102D} for ultramagnetized white dwarfs, 
see below Eq.~(\ref{eq:Mmax}), neglects all the above macro and micro physical aspects relevant for the self-consistent 
description of the structure and assessment of stability of these objects. When accounted for, they lead to the conclusion 
that the existence of such ultramagnetized white dwarfs in nature is very unlikely due to violation of minimal requests of 
stability. Indeed, all these ignored effects make improbable that a white dwarf could reach such a hypothetical extreme 
state either in single or binary evolution.

Therefore, the canonical Chandrasekhar mass limit of white dwarfs has to be still applied and consequently, ultramagnetized white
dwarfs cannot be used as progenitors of superluminous supernovae.

%%%%%%%%%%%%%%%%%%%%%%%%%%%%%%%%%%%%%%%%%%%%%%%%%%%%%%%%%%%%%%%%%%%%%%%%%%%%%%%%%%%%%%%%%%%%%%%%%%
%%%%%%%%%%%%%%%%%%%%%%%%%%%%%%%%%%%%%%%%%%%%%%%%%%%%%%%%%%%%%%%%%%%%%%%%%%%%%%%%%%%%%%%%%%%%%%%%%%
\section{Ultramagnetized white dwarfs}\label{sec:2}
%%%%%%%%%%%%%%%%%%%%%%%%%%%%%%%%%%%%%%%%%%%%%%%%%%%%%%%%%%%%%%%%%%%%%%%%%%%%%%%%%%%%%%%%%%%%%%%%%%
%%%%%%%%%%%%%%%%%%%%%%%%%%%%%%%%%%%%%%%%%%%%%%%%%%%%%%%%%%%%%%%%%%%%%%%%%%%%%%%%%%%%%%%%%%%%%%%%%%

In a recent work, \citet{2013PhRvL.110g1102D} studied
the effects of extreme magnetic field in the mass and radius of white dwarfs. They showed that the EOS of a degenerate
electron gas in presence of a magnetic field $B$ directed along the $z$-axis, in the limit $B\to\infty$ when all electrons 
are constrained to the lowest Landau level, obeys a polytrope-like
\begin{equation}\label{eq:Km}
P=K_m\rho^2,\qquad K_m=\frac{m_ec^2\pi^2\lambda_e^3}{(\mu_e m_H)^2B_D},
\end{equation}
with $\lambda_e$ the electron Compton wavelength, and $B_D=B/B_c$ the magnetic field in units of the critical
field $B_c=m^2_e c^3/(e \hbar)=4.41\times10^{13}$~G. For obtaining the above expression, in \citep{2013PhRvL.110g1102D}
the density of the system was assumed to be given by $\rho=\mu_e m_H n_e$, so determined only by the nuclei component, where
$n_e$ is the electron number density.

Then, Lane-Emden solution of Newtonian self-gravitating polytropes of index $n=1$ was used to obtain the mass of an ultramagnetized white dwarf
\begin{equation}\label{eq:M}
 M=4\pi^2\rho_c\left(\frac{K_m}{2\pi G}\right)^{3/2},
\end{equation}
and the corresponding radius
\begin{equation}\label{eq:R}
 R=\sqrt{\frac{\pi K_m}{2G}},
\end{equation}
where $\rho_c$ is the central density. 

In the present limit of one Landau level with high electron Fermi energies $E^F_e$, $E^F_e=E^F_{\rm max}\gg m_e c^2$, with 
\begin{equation}\label{eq:Efmax}
E^F_{\rm max}=m_e c^2\sqrt{1+2B_D}\approx m_e c^2\sqrt{2 B_D}
\end{equation}
the maximum possible value of $E^F_e$, $\rho_c$ becomes 
\begin{equation}\label{eq:rhoc}
 \rho_c= \frac{\pi M}{4R^3} = \frac{\mu_e m_H}{\sqrt{2} \pi^2 \lambda_e^3}B_D^{3/2}.
\end{equation}

Introducing Eq.~(\ref{eq:rhoc}) into Eq.~(\ref{eq:M}), \citet{2013PhRvL.110g1102D} obtained the
mass limit of ultramagnetized white dwarfs 
\begin{equation}\label{eq:Mmax}
M_{\rm max}=\pi^{3/2} \frac{m^3_{\rm Pl}}{(\mu_e m_H)^2}\approx 2.58~M_\odot,
\end{equation}
when $\rho_c\to \infty$ and $R\to 0$. This upper bound is larger than the canonical Chandrasekhar limit given by Eq.~(\ref{eq:chandralimit}).

\begin{figure}[!hbtp]
\centering
\includegraphics[width=0.7\hsize,clip]{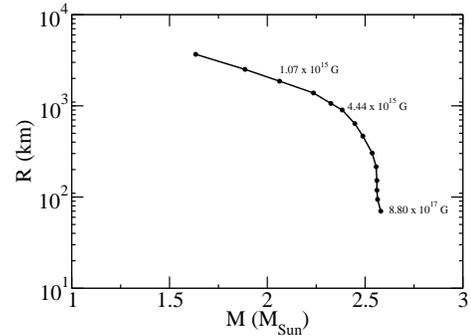}
\caption{Mass-radius relation of magnetized white dwarfs - the curve represents the evolutionary track of
the white dwarf with the increase of the uniform magnetic field inside the star obtained in \citep{2013PhRvL.110g1102D}.}\label{figure1}
\end{figure}

We reproduce in Fig.~\ref{figure1} the evolutionary track of the white dwarf proposed in \citep{2013PhRvL.110g1102D}. 
The magnetic field along the curve is increasing as a consequence of accretion of matter onto the star. It can be seen
in the plot how the star reaches the maximum mass limit (\ref{eq:Mmax}) while reducing its radius.

Already at this point it is possible to identify some of the assumptions in the model of \citet{2013PhRvL.110g1102D} 
that led to the above results, and which as we show below are unjustified, invalidating their final conclusions.
1) The EOS assumed in the limit of very intense magnetic fields, $B\to \infty$; 2) a uniform magnetic field is adopted;
3) the huge magnetic fields and the obtained mass-radius relation explicitly violate even the absolute upper limit to the magnetic field
imposed by the Virial theorem; 4) dynamical instabilities due to quadrupole deformation are not taken into account either; 5) spherical 
symmetry is assumed for all values of the magnetic field; 6) the role of the magnetic field in the hydrostatic equilibrium equations is
neglected; 7) general relativistic effects are ignored even if the final configuration is almost as compact as a neutron star and the 
magnetic energy is larger than the matter energy-density; 8) microphysical effects such as inverse $\beta$ decay and pycnonuclear fusion 
reactions, important in a regime where the electrons are highly relativistic, $E^F_e \gg m_e c^2$, are neglected; and 9) the magnetic 
field, the density, and the electron Fermi energy are assumed to increase with time inside the star as a consequence of a continuous 
accretion process onto the white dwarf.

%%%%%%%%%%%%%%%%%%%%%%%%%%%%%%%%%%%%%%%%%%%%%%%%%%%%%%%%%%%%%%%%%%%%%%%%%%%%%%%%%%%%%%%%%%%%%%%%%%
%%%%%%%%%%%%%%%%%%%%%%%%%%%%%%%%%%%%%%%%%%%%%%%%%%%%%%%%%%%%%%%%%%%%%%%%%%%%%%%%%%%%%%%%%%%%%%%%%%
\section{Equation of state and virial theorem violation}\label{sec:3}
%%%%%%%%%%%%%%%%%%%%%%%%%%%%%%%%%%%%%%%%%%%%%%%%%%%%%%%%%%%%%%%%%%%%%%%%%%%%%%%%%%%%%%%%%%%%%%%%%%
%%%%%%%%%%%%%%%%%%%%%%%%%%%%%%%%%%%%%%%%%%%%%%%%%%%%%%%%%%%%%%%%%%%%%%%%%%%%%%%%%%%%%%%%%%%%%%%%%%

Being much lighter, the electrons in the white dwarf interior are more easily disturbed by a magnetic field than the ions. Eventually, 
the electron gas might become quantized in Landau levels, providing the magnetic field is larger than the critical
field $B_c$. However, for ``moderate'' values of the field, i.e. $B\sim B_c$, the EOS deviates still very little 
from the unmagnetized one. Thus, appreciable effects are seen only when the electrons occupy only the lower Landau levels, which
is possible for $B_D\sim [E^F_{\rm max}/(m_e c^2)]^2$. Since the electrons in massive white dwarfs are ultrarelativistic with Fermi 
energies $E^F_e\gtrsim 10~m_e c^2$ \citep{2011PhRvD..84h4007R}, it implies the necessity of magnetic 
fields $B_D\gtrsim 10^2$ ($B\gtrsim 4\times 10^{15}$~G), in order to have non-negligible magnetic field effects.
It can be checked from the virial theorem that such large magnetic fields 
cannot develop in the interior of the white dwarf since they violate the absolute upper bound imposed by the virial theorem applied 
to a white dwarf which is approaching the Chandrasekhar mass limit.

The limiting field can be computed following the argument by \citet{1953ApJ...118..116C} in their seminal work.
There exists a magnetic field limit, $B_{\rm max}$, above which an equilibrium configuration is impossible because the electromagnetic 
energy, $W_B$, exceeds the gravitational energy, $W_G$, therefore becoming gravitationally unbound. If one includes the forces derived 
from the magnetic field,
one can write the virial scalar relation for an equilibrium configuration 
as \citep{1953ApJ...118..116C}
\begin{equation}\label{eq:virial_eq}
 3 \Pi+W_B+W_G=0,
\end{equation}
where $\Pi=\int P d V$, with $P$ the pressure of the system, $W_B$ the positive magnetic energy, and $W_G$ the negative gravitational
potential energy. The quantity $\Pi$ satisfies $\Pi=(\gamma-1)U$ for a polytrope, $P=K \rho^\gamma$, where $U$ is the total kinetic energy 
of particles. Since the total energy of the configuration can be written as $E=U+W_B+W_G$, then one can eliminate $U$ from Eq.~(\ref{eq:virial_eq})
to obtain $E=-[(\gamma-4/3)/(\gamma-1)](\lvert{W_G}\lvert-W_B)$, and therefore the necessary condition for the stability of the star, $E<0$, is given by
\begin{equation}
 (3\gamma -4)\lvert W_G\lvert\left(1-\frac{W_B}{\lvert W_G\lvert}\right)>0.
\end{equation}

From this expression we can recover, in absence of magnetic field ($W_B = 0$), the known condition for bound 
unmagnetized polytropes $\gamma<4/3$, or $n<3$ in terms of the polytrope index $n$ defined by $\gamma=1+1/n$. The presence of a magnetic
field weakens the stability, and no matter the value of $\gamma$, the star becomes gravitationally unbound when the magnetic energy
exceeds the gravitational one; i.e. $W_B>\lvert W_G\lvert$. This condition clearly implies an upper bound for the magnetic field, 
obtained for $W_B=\lvert W_G\lvert$. In order to determine such limit
we first obtain an expression for the magnetic energy of the star, which considering a constant magnetic field can be written as
\begin{equation}
 W_B=\frac{B^2}{8\pi} \frac{4\pi R^3}{3}=\frac{B^2R^3}{6}.
\end{equation}

As we discussed above, the EOS used by \citet{2013PhRvL.110g1102D} adopts a polytrope-like form with $\gamma=2$ or $n=1$ under extreme magnetic fields, 
such that only one Landau level is populated and $E_F\gg m_ec^2$. Thus, the gravitational energy density
of the spherical star configuration is \citep{1983bhwd.book.....S},
\begin{equation}
 W_G=-\frac{3}{5-n}\frac{GM^2}{R}=-\frac{3}{4}\frac{GM^2}{R},
\end{equation}

where $M$ and $R$ are the mass and star radius, respectively, and $G$ is the Newton gravitational constant. Using
the above expressions, and expressing $M$ and $R$ in units of solar mass and solar radius, we find 
that the maximum value of magnetic field $B_{\rm max}$ is given by
\begin{equation}\label{eq:Bmax}
 B_{\rm max}=2.24\times10^8\frac{M}{M_\odot}\left(\frac{R_\odot}{R}\right)^2~{\rm G}.
\end{equation}

In the case of a Chandrasekhar white dwarf with the maximum mass $M=1.44M_\odot$ and a radius of 3000~km, consistent with
the recent calculation of massive white dwarfs \citep{2013ApJ...762..117B}, we obtain $B_{\rm max}\sim 1.7\times 10^{13}$ G. 
This value is clearly lower than the critical field $B_c=4.4\times 10^{13}$ G.

\begin{figure}[!hbtp]
\begin{center}
\includegraphics[width=0.7\hsize,clip]{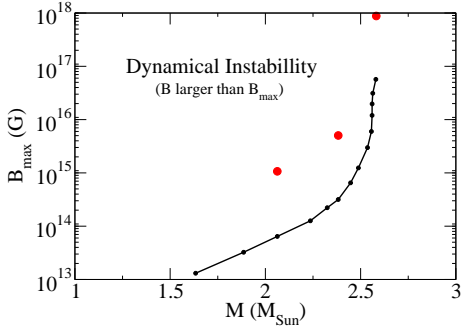}\\
\caption{(Color online) Maximum magnetic field $B_ {\rm max}$ as a function of the star mass.
We show (red dots) the three values of the magnetic field of Table 1
that are above the $B_ {\rm max}$ line, in the dynamical instability region.}\label{figure2}
\end{center}
\end{figure}

In order to quantify how strong is the violation of the virial theorem produced by the magnetic fields used in \citep{2013PhRvL.110g1102D},
we choose three star configurations whose values of $M$ and $R$ lie in the region of high mass configuration, $M>2M_\odot$ (see red points 
in Fig.~\ref{figure2}). Using the approximation of Eq.~(\ref{eq:rhoc}), we obtain the corresponding constant magnetic field $B$ of these stars
configurations. We compare these values of $B$ with the maximum value, $B_{\rm max}$, allowed by the virial theorem (\ref{eq:Bmax}). 
In Fig.~\ref{figure2} we present the virial limit $B_{\rm max}$ as a function of the star mass obtained by Eq.~(\ref{eq:Bmax})
using the values of mass and radius shown in Fig.~\ref{figure1}. In this figure we show that such extreme magnetic fields with $B>B_{\rm max}$ 
and the magnetized white dwarfs of Table~\ref{ta2}
are in the instability region, violating the virial theorem. In Table~\ref{ta2} we show also for these configurations the magnetic 
energy $W_B$, and the magnitude of the gravitational energy $\lvert W_G\lvert$. These results indicate that
the magnetic field obtained in \citep{2013PhRvL.110g1102D}, are at least one order of magnitude larger than the maximum magnetic field
allowed, $B_{\rm max}$. As a consequence, for the three star configurations, $W_B/\lvert W_G\lvert \sim250$ well above the stability condition which
requires $W_B/\lvert W_G\lvert\sim1$. Thus, these white dwarf are unstable and unbound.

The repulsive magnetic force due to a possible variable magnetic field, as discussed in \citep{2007IJMPD..16..489M}, was not considered.
Furthermore, a uniform magnetic field in the $z$-direction inside the star, yields a dipole external field \citep{1953ApJ...118..116C}. In this
case, even if the magnetic fields are continuous at the star surface, their derivatives are not, producing a repulsive magnetic force at the
surface. This force will push against the attractive gravitational force such that for a large magnetic field, the magnetic force will 
overcome the gravitational one, destabilizing the star. This physical situation is exactly the same expressed in the virial theorem condition
for the star stability ($W_B<\lvert W_G\lvert$) discussed above. 
\citet{1968ApJ...153..797O} analyzed the effect of magnetic fields in white dwarfs, and concluded that they lead to
stars with larger masses but also larger radii. One of the main consequences of the increasing magnetic field is that even a small ratio of magnetic to gravitational energy will produce an appreciable increase in the radii of magnetized white dwarfs. Consequently, it leads to a reduction of the central density,
even for small mass changes. These conclusions were also confirmed in \citep{2000ApJ...530..949S}, where the effect of magnetic fields 
in the mass-radius relation for magnetic white dwarfs were also investigated. Thus, the very compact magnetized white dwarf configuration 
obtained in \citep{2013PhRvL.110g1102D}, in which large magnetic field implies large mass and {\it small} radius, are possible only
because the effect of the repulsive magnetic force (Lorentz force) has not been properly considered.
Since in \citep{2013PhRvL.110g1102D} it is considered the influence of a very large constant magnetic fields in the star mass and radius, 
assuming values for the magnetic field larger than the above limits, we conclude that these extremely magnetized white dwarfs must be unstable 
and unbound. The limiting magnetic field values $B_ {\rm max}$ shown in Fig.~\ref{figure2} and Table~\ref{ta2} are clearly obtained with the radii given in \citep{2013PhRvL.110g1102D}, which are much smaller than the self-consistent solution to the equilibrium equations would give. Since the maximum magnetic field depends on $R^{-2}$, see Eq.~(\ref{eq:Bmax}), the real maximum possible field would actually be smaller than the one computed here.

Indeed, it is worth to notice that the Eq.~(\ref{eq:Bmax}) can be also expressed as 
a limit to the magnetic flux: $\Phi_{\rm max}\sim B_{\rm max} R^2 \approx 1.1\times 10^{30} (M/M_\odot)$~G~cm$^2$. 
For the hypothetic new maximum mass (\ref{eq:Mmax}), $M=2.58\,M_\odot$, this
maximum magnetic flux is $\Phi_{\rm max}\approx 2.8\times 10^{30}$~G~cm$^2$. It 
is interesting that Eqs.~(\ref{eq:Km}--\ref{eq:rhoc}) imply magnetic flux-freezing,
namely a constant value of the magnetic 
flux, $\Phi_{\rm frozen}/B_c\sim B_D R^2=\pi^3 (\hbar c/G)(\mu_e m_H)^{-2}\lambda_e^2\approx 2\times 10^{18}$~cm$^2$, or $\Phi_{\rm frozen}\sim B R^2\approx 8.74 \times 10^{31}$~G~cm$^2$. 
This constant value highly overcomes the above maximum possible magnetic flux, $\Phi_{\rm max}$, 
which shows in a different way the violation of the stability limit imposed by the virial theorem by
the solution presented by \citet{2013PhRvL.110g1102D}.

%%%%%%%%%%%%%%%%%%%%%%%%%%%%%%%%%%%%%%%%%%%%%%%%%%%%%%%%%%%%%%%%%%%%%%%%%%%%%%%%%%%%%%%%%%%%%%%%%%
%%%%%%%%%%%%%%%%%%%%%%%%%%%%%%%%%%%%%%%%%%%%%%%%%%%%%%%%%%%%%%%%%%%%%%%%%%%%%%%%%%%%%%%%%%%%%%%%%%
\section{Breaking of spherical symmetry and quadrupole instability}\label{sec:4}
%%%%%%%%%%%%%%%%%%%%%%%%%%%%%%%%%%%%%%%%%%%%%%%%%%%%%%%%%%%%%%%%%%%%%%%%%%%%%%%%%%%%%%%%%%%%%%%%%%
%%%%%%%%%%%%%%%%%%%%%%%%%%%%%%%%%%%%%%%%%%%%%%%%%%%%%%%%%%%%%%%%%%%%%%%%%%%%%%%%%%%%%%%%%%%%%%%%%%

It was shown by \citet{1953ApJ...118..116C} that the figure of equilibrium of an incompressible fluid sphere with an 
internal uniform magnetic field that matches an external dipole field, is not represented by a sphere. The star becomes oblate by contracting 
along the axis of symmetry, namely along the direction of the magnetic field. Thus, we consider the fluid sphere to be
deformed in such a way that the equation of the bounding surface is given by
\begin{equation}
 r(\mu)=R+\epsilon P_l(\mu),
\end{equation}
where $\mu=\cos\theta$, with $\theta$ the polar angle, and $P_l(\mu)$ denotes the Legendre polynomial of order $l$. It 
is easy to see that the deviation from the spherical configuration is given by the term $P_l(\mu)$, thus in \citep{1953ApJ...118..116C},
such a perturbation was called 
\textquotedblleft $P_l$- deformation\textquotedblright.
They have also concluded that the term with $l=2$ contributes to the
corresponding change in the internal magnetic energy density (for all other
values of $l$, the change in the magnetic energy is of 
the second order in $\epsilon$).
The quantity $\epsilon$ satisfies $\epsilon\ll R$ and measures the deviations from 
a spherical configuration. The polar and equatorial radii are $R_p=R+\epsilon P_l(1)$ and $R_{\rm eq}=R+\epsilon P_l(0)$ 
respectively, thus $\epsilon=-(2/3)(R_{\rm eq}-R_p)$ and therefore $\epsilon/R=-(2/3)(R_{\rm eq}-R_p)/R$, for the axisymmetric deformed 
configuration with $l=2$.

It was shown by \citet{1953ApJ...118..116C} that such an axisymmetrically deformed object is favorable energetically with respect to the 
spherical star. Thus, the star becomes unstable and proceeds to collapse along the magnetic field axis, turning into an oblate spheroidal 
shape with $\epsilon < 0$. The contraction continues until the configuration reaches a value of $\epsilon/R$ given by
\begin{equation}
 \frac{\epsilon}{R}=-\frac{35}{24}\frac{B^2 R^4}{GM^2}.
\end{equation}

Using the expression for $B_{\rm max}$ given by Eq.~(\ref{eq:Bmax}), one obtains 
\begin{equation}
 \frac{\epsilon}{R}=-\frac{315}{384}\left(\frac{B}{B_{\rm max}}\right)^2\simeq -0.8\left(\frac{B}{B_{\rm max}}\right)^2.
\end{equation}
Therefore, when the internal magnetic field is close to the
limit set by the virial theorem, the star
deviates to a highly oblate shape.

We show in the last column of Table~\ref{ta2}, the \textquotedblleft $P_l$- deformation\textquotedblright, $\epsilon/R$, calculated
for three configurations discussed before. The results show that $\lvert
\epsilon/R\lvert\gtrsim2\times 10^2$, which implies that the star has a highly oblate shape and thus the spherical symmetry is strongly
broken. Therefore, in order to account for the deformation caused by the presence of a magnetic field, a more consistent calculation 
considering cylindrical symmetry, see e.g.~\citep{1953ApJ...118..116C,1968ApJ...153..797O}, is mandatory. 
It is worth to mention 
that if we consider the quantum nature of the EOS of a Fermi gas subjected at fields $B\gg B_c$, the actual shape 
of equilibrium is defined by a the total (matter+field) pressures parallel and perpendicular
to the magnetic field that are different, and vanish at the star surface \citep{PRL2000Aurora,EPJ2003Aurora,IJMPD2008Aurora,PRD2012Debora}.

%%%%%%%%
\begin{table*}[!t]
\caption{Table with the mass-radius configurations of magnetized white dwarfs of \citet{2013PhRvL.110g1102D} with the correspondent
magnetic field $B$, the maximum virial magnetic
field $B_{\rm max}$, magnetic energy $W_B$ and gravitational $W_G$, the ratio of them $W_B/\lvert W_G\lvert$, the magnetic field contribution to the 
total energy density $\rho_B$,
and the values of eccentricity in units of the spherical star radius $\epsilon/R$.}
\newcommand{\cc}[1]{\multicolumn{1.5}{c}{#1}}
\renewcommand{\tabcolsep}{0.05pc} % enlarge column spacing
\renewcommand{\arraystretch}{1.2} % enlarge line spacing
\begin{ruledtabular}
%\begin{tabular}{|l|p{3cm}|c|}
{\begin{tabular}{@{}cccccccccccccccccc@{}} 
&\footnotesize $M$ $\rm(M_\odot)$ &\footnotesize$R$ $\rm(km)$ &\footnotesize$B$ $\rm (G)$ &\footnotesize$B_{\rm max}$ $\rm(G)$ &\footnotesize$W_B$ $(\times10^{51}\rm erg)$ &\footnotesize$\lvert W_G\lvert$ $(\times10^{51}\rm erg)$    &\footnotesize$W_B/\lvert W_G\lvert$ $\rm$   &\footnotesize$\rho_B$ (~g~cm$^{-3}$) &\footnotesize$\epsilon/R$\\
\tableline
&2.58&              7.02$\times10^1$&    8.80$\times10^{17}$&   5.70$\times10^{16}$&         4.43$\times10^{4}$&      1.88$\times10^{2}$&        235&      3.40$\times10^{13}$&       -195.14\\ 
&2.38&              9.60$\times10^2$&    4.44$\times10^{15}$&   2.81$\times10^{14}$&         2.90$\times10^{3}$&      1.17$\times10^{1}$&         248&     8.71$\times10^{8}$&        -204.16  \\
&2.06&              1.86$\times10^3$&    1.07$\times10^{15}$&   6.49$\times10^{13}$&         1.23$\times10^{3}$&      4.52&         273&      5.10$\times10^{7}$&        -223.03 \\  
\end{tabular} \label{ta2}}
\end{ruledtabular}
\end{table*}

%%%%%%%%%%%%%%%%%%%%%%%%%%%%%%%%%%%%%%%%%%%%%%%%%%%%%%%%%%%%%%%%%%%%%%%%%%%%%%%%%%%%%%%%%%%%%%%%%%
%%%%%%%%%%%%%%%%%%%%%%%%%%%%%%%%%%%%%%%%%%%%%%%%%%%%%%%%%%%%%%%%%%%%%%%%%%%%%%%%%%%%%%%%%%%%%%%%%%
\section{Microscopic instabilities}\label{sec:5}
%%%%%%%%%%%%%%%%%%%%%%%%%%%%%%%%%%%%%%%%%%%%%%%%%%%%%%%%%%%%%%%%%%%%%%%%%%%%%%%%%%%%%%%%%%%%%%%%%%
%%%%%%%%%%%%%%%%%%%%%%%%%%%%%%%%%%%%%%%%%%%%%%%%%%%%%%%%%%%%%%%%%%%%%%%%%%%%%%%%%%%%%%%%%%%%%%%%%%

It is known that at sufficiently high densities in the interior of white dwarfs, the inverse $\beta$ decay or electron capture process becomes
energetically favorable, and therefore a nucleus $(Z,A)$ transforms into a different nucleus $(Z-1,A)$ by capturing energetic electrons.
Such a process destabilizes the star since the electrons are the main responsible for the pressure in a white dwarf 
\citep{harrison1958,1965gtgc.book.....H,1983bhwd.book.....S}. The process sets in when the electron chemical potential
reaches the threshold energy, $\epsilon^\beta_Z$, given by the difference of the nuclear binding energy between the 
initial and final nucleus. For helium, carbon, oxygen, and iron, $\epsilon^\beta_Z$ is approximately
20.6, 13.4, 10.4, and 3.7~MeV respectively \citep[see, e.g.,][]{tuli2011}. For unmagnetized general 
relativistic white dwarfs, this occurs at a critical density
$\rho^\beta_{\rm crit}\approx 1.4\times 10^{11}$, $4.0\times 10^{10}$, $1.9\times 10^{10}$, 
and $1.2\times 10^9$~g~cm$^{-3}$, respectively for the same chemical compositions \citep[see Table II in][]{2011PhRvD..84h4007R}.

This instability was recently analyzed by \citet{2013PhRvD..88h1301C} for the ultramagnetized white 
dwarfs discussed in \citep{2013PhRvL.110g1102D}. Using
Eq.~(\ref{eq:Efmax}), it can be seen that the electron capture process limits the
magnetic field to values lower than \citep[see, e.g.,][]{2013PhRvD..88h1301C}
\begin{equation}\label{eq:Bmaxbeta}
B^\beta_D= \frac{1}{2}\left(\frac{\epsilon^\beta_Z}{m_e c^2}\right)^2\approx 812.6,~342.3,~207.9,~26.2,
\end{equation}
or $B\approx 3.6\times 10^{16}$, $1.5\times 10^{16}$, $9.1\times 10^{15}$, and $1.1\times 10^{15}$~G, where we have used the previously
mentioned values of $\epsilon^\beta_Z$ for helium, carbon, oxygen, and iron, respectively. We can show that the electron
capture in the configurations of \citet{2013PhRvL.110g1102D} occurs even at critical central densities lower
than in the unmagnetized case. Indeed, by introducing the limiting values of Eq.~(\ref{eq:Bmaxbeta}) into
Eq.~(\ref{eq:rhoc}), we obtain the values of the critical
densities $\rho^\beta_{\rm crit}\approx 9.6\times 10^{10}$, $2.6\times 10^{10}$, $1.2\times 10^{10}$, and $6.0\times 10^{8}$~g~cm$^{-3}$, 
respectively 
for helium, carbon, oxygen, and iron. These densities are much smaller than the ones of the massive ultramagnetized white dwarfs considered
by \citet{2013PhRvL.110g1102D}: the configurations approaching the maximum mass (\ref{eq:Mmax}) have
magnetic fields $B_D \gtrsim 10^4$ ($B\gtrsim 4\times 10^{17}$~G) and therefore central densities
$\rho_c\gtrsim 4\times 10^{12}$~g~cm$^{-3}$. At such high densities, higher than 
the neutron drip value ($\rho_{\rm drip}\approx 4.3\times 10^{11}$~g~cm$^{-3}$), the less bound neutrons in nuclei start to drip 
out forming a free Fermi gas \citep{baym71}. The neutron drip process then starts when $\rho_c=\rho_{\rm drip}$ where $\rho_c$ is given 
by Eq.~(\ref{eq:rhoc}). For a carbon composition, it occurs for a magnetic field $B_D\approx 531$, or $B\approx 2.3\times 10^{16}$~G 
\citep[see, e.g.,][]{2013PhRvD..88h1301C}. It is important to clarify that extremely large magnetic fields ($>10^{17}$~G)
are needed to modify the neutron drip value appreciable, and we refer the reader to \cite{PhysRevC.86.055804} for an 
analysis of the influence of strong mangetic fields on the precise value of the neutron drip density and pressure.

As discussed by \citet{2013PhRvD..88h1301C}, pycnonuclear fusion reactions might establish a 
more stringent limit with respect to the inverse $\beta$ decay in an ultramagnetized white dwarf. 
Carbon fusion leads to $^{24}$Mg, which undergoes electron capture, thus inverse $\beta$ decay instability,
at a density of approximately $\rho^{\beta}_{\rm crit,Mg}\approx 3\times 10^9$~g~cm$^{-3}$. Therefore, if C+C fusion occurs at rates highly
enough at densities lower than $\rho^{\beta}_{\rm crit,Mg}$ to produce appreciable amounts of $^{24}$Mg in times shorter than a Hubble time, then this 
process imposes a more tight constraint to the density of the white dwarf. Based on the up-to-date astrophysical S-factors computed in 
\citep{2005PhRvC..72b5806G}, we recently computed in \citep{2013ApJ...762..117B} the pycnonuclear carbon fusion rates in white dwarfs. We 
found for instance that, C+C fusion occurs at a timescale of 0.1~Myr at
a density $\rho^{\rm C+C}_{\rm pyc}\approx 1.6\times 10^{10}$~g~cm$^{-3}$. 
Since $\rho^{\rm C+C}_{\rm pyc}<\rho^{\beta}_{\rm crit,C}\approx 2.6\times 10^{10}$~g~cm$^{-3}$,
this implies that C+C pycnonuclear fusion does limit further the magnetic field strength with respect
to the inverse $\beta$ decay instability of carbon. Indeed, using Eq.~(\ref{eq:rhoc}), we obtain that
such a density is reached for a magnetic field
$B^{\rm C+C}_{D,\rm pyc}\approx 246.6$, or $B^{\rm C+C}_{\rm pyc}\approx 1.1\times 10^{16}$~G,
a value lower than $B^{\beta,{\rm C}}_D\approx 342.3$ or $B^\beta_{\rm C}\approx 1.5\times 10^{16}$~G. Longer reaction 
times implies lower densities and thus lower magnetic fields.

It is important to note that the above limits to the magnetic field are estimated assuming that
the density of the system is given by Eq.~(\ref{eq:rhoc}); however more realistic estimates of these 
limiting fields have to account for the contribution of the magnetic field to the mass-energy density 
(see below in section \ref{sec:6}) and the self-consistent value of the electron density accounting for
the real number of Landau levels populated, which will be higher than one. The above microscopic limits
to the magnetic field are anyway higher than the maximal values allowed 
by the virial theorem. Therefore, the macroscopic dynamical instabilities appear to set in before 
both electron captures and pycnonuclear reactions.

%%%%%%%%%%%%%%%%%%%%%%%%%%%%%%%%%%%%%%%%%%%%%%%%%%%%%%%%%%%%%%%%%%%%%%%%%%%%%%%%%%%%%%%%%%%%%%%%%%
%%%%%%%%%%%%%%%%%%%%%%%%%%%%%%%%%%%%%%%%%%%%%%%%%%%%%%%%%%%%%%%%%%%%%%%%%%%%%%%%%%%%%%%%%%%%%%%%%%
\section{General relativistic effects}\label{sec:6}
%%%%%%%%%%%%%%%%%%%%%%%%%%%%%%%%%%%%%%%%%%%%%%%%%%%%%%%%%%%%%%%%%%%%%%%%%%%%%%%%%%%%%%%%%%%%%%%%%%
%%%%%%%%%%%%%%%%%%%%%%%%%%%%%%%%%%%%%%%%%%%%%%%%%%%%%%%%%%%%%%%%%%%%%%%%%%%%%%%%%%%%%%%%%%%%%%%%%%

We now turn to show that for ultra high magnetic fields as the ones considered by \citet{2013PhRvL.110g1102D}, general
relativistic effects are relevant; therefore a Newtonian treatment of the equations of equilibrium is not appropriate. 
First we can calculate the contribution of an ultra high magnetic field, as the ones considered in
\citep{2013PhRvL.110g1102D}, to the total energy-density. For the  maximum white dwarf mass in \citep{2013PhRvL.110g1102D}, Eq.~(\ref{eq:Mmax}),
which is obtained for a magnetic field $B\approx 10^{18}$~G, the magnetic field contribution to the 
total energy-density is $\rho_B\approx B^2/(8\pi c^2)\approx 4.4\times 10^{13}$~g~cm$^{-3}$. 
This value is indeed larger than the matter density of the configuration and cannot be therefore neglected in the energy-momentum tensor of the system.
However, as we have shown such a large magnetic 
fields cannot be reached in the star; thus the real configurations of equilibrium likely have a magnetic field energy-density 
much smaller than the matter energy-density, implying that the unmagnetized maximum mass, the Chandrasekhar mass $M_{\rm Ch}\approx 1.44~M_\odot$,
still applies.

On the other hand, when the maximum mass (\ref{eq:Mmax}) is approached for magnetic fields $B_D \gtrsim 10^4$, the
central density of the system, as given by Eq.~(\ref{eq:rhoc}), is $\rho_c\gtrsim 4\times 10^{12}$~g~cm$^{-3}$. In particular, the maximum mass configuration would have a radius $R\approx 70$~km and therefore a central density $\rho_c\approx 1.2\times 10^{13}$~g~cm$^{-3}$, only one order of magnitude 
less than the nuclear saturation density. These values imply that the mass, radius, and density of the ultramagnetized objects 
considered by \citet{2013PhRvL.110g1102D} are much more similar to the parameters of neutron star rather than to the ones of a white
dwarf. Therefore, it is natural to ask whether the compactness of the star is such to require a full general 
relativistic treatment. For the above star parameters close to the maximum mass configuration, it is obtained a compactness $G M/(c^2 R)\approx 0.05$, 
a value in clear contrast with a Newtonian treatment of the equilibrium equations.

In this line, our previous results~\citep{2011PhRvD..84h4007R} become relevant. We found there that, in the case of carbon white
dwarfs, general relativistic instability sets in at a density $\rho_{\rm crit}\approx 2\times 10^{10}$~g~cm$^{-3}$, prior to the
inverse $\beta$ decay instability. Such a density is much lower than the densities of the ultramagnetized white dwarfs of \citet{2013PhRvL.110g1102D}.

%%%%%%%%%%%%%%%%%%%%%%%%%%%%%%%%%%%%%%%%%%%%%%%%%%%%%%%%%%%%%%%%%%%%%%%%%%%%%%%%%%%%%%%%%%%%%%%%%%
%%%%%%%%%%%%%%%%%%%%%%%%%%%%%%%%%%%%%%%%%%%%%%%%%%%%%%%%%%%%%%%%%%%%%%%%%%%%%%%%%%%%%%%%%%%%%%%%%%
\section{Evolutionary path}\label{sec:7}
%%%%%%%%%%%%%%%%%%%%%%%%%%%%%%%%%%%%%%%%%%%%%%%%%%%%%%%%%%%%%%%%%%%%%%%%%%%%%%%%%%%%%%%%%%%%%%%%%%
%%%%%%%%%%%%%%%%%%%%%%%%%%%%%%%%%%%%%%%%%%%%%%%%%%%%%%%%%%%%%%%%%%%%%%%%%%%%%%%%%%%%%%%%%%%%%%%%%%

As a possible mechanism of formation of ultramagnetized white dwarfs, it was proposed
in \citep{2013PhRvL.110g1102D} and further extended in \citep{2013ApJ...767L..14D}
the traditional idea that the star by accretion could increase continuously its central
density and its magnetic field owing to magnetic flux conservation. However, it is 
unlikely that such an accretion could bring the white dwarf to such extreme regimes without 
passing through all the instability channels analyzed in this work. It can be shown that
the magnetic field, by flux conservation, cannot increase by orders of magnitude during the
accretion process if we account for the stability limits and the realistic structure of the
white dwarf. Flux conservation implies, for a uniform magnetic field as assumed 
by \citet{2013PhRvL.110g1102D}, $B_f/B_0 =(R_0/R_f)^2$ where `0' and `$f$' stands 
for initial and final values. It is known that in the Newtonian treatment the critical 
mass is reached at infinite densities, so when the radius tends to zero, causing an
unphysical large increase of the above magnetic field when approaching the critical
mass value. Therefore, taking into due account general relativistic and microscopic
instabilities leading to a finite critical density and radius for the critical mass
configuration are essential in this computation. For this purpose we use the mass-radius 
relation obtained by \citet{2011PhRvD..84h4007R}. If we start an accretion process on a 
carbon white dwarf with initial mass $M_0\sim 1\,M_\odot$ ($R_0\approx 5587.43$~km), typical
of high magnetic field white dwarf population \citep[see][for details]{2005MNRAS.361.1131F}, 
we obtain that the magnetic field increases only a factor $B_f/B_0\approx 28$ up to the final
mass $M_f=M_{\rm crit}\approx 1.39\,M_\odot$ ($R_f\approx 1051.44$~km). Indeed, the magnetic 
flux is $\Phi\sim B_0 R_0^2\approx 3.1\times 10^{25}(B_0/10^8)$~G~cm$^2$, to be contrasted with 
much higher value of the frozen value $\Phi_{\rm frozen}\approx 8.7\times 10^{31}$~G~cm$^2$, inferred 
at the end section \ref{sec:3} for the maximum mass solution of \citet{2013PhRvL.110g1102D}. This implies that,
most likely the accretion will lead in due time to the triggering of the white dwarf gravitational collapse to
a neutron star, or to an ordinary type Ia supernovae, prior to reach a stage where the magnetic field causes
appreciate changes to the EOS and to the structure of the star. One could think that the white dwarf has already
a huge magnetic field ($\gtrsim 10^{15}$~G) before starting the accretion process. However, as we have shown in 
section \ref{sec:3}, the virial theorem imposes a limiting magnetic
flux $\Phi_{\rm max}\approx 1.1\times 10^{30} \sqrt{4/(5-n)}\,(M/M_\odot)$~G~cm$^2$, where $n$ is
the polytropic index, which anyway limits the magnetic field of the initial configuration to lower values. 
In addition, huge seed magnetic fields in the interior of a solar mass white dwarf appear to be in
contradiction with observations since the unmagnetized mass-radius relation reproduces with 
appreciable accuracy the observational data \citep[see, e.g.,][]{1997A&A...325.1055V,1998ApJ...494..759P}.

%%%%%%%%%%%%%%%%%%%%%%%%%%%%%%%%%%%%%%%%%%%%%%%%%%%%%% %%%%%%%%%%%%%%%%%%%%%%%%%%%%%%%%%%%%%%%%%%%%
%%%%%%%%%%%%%%%%%%%%%%%%%%%%%%%%%%%%%%%%%%%%%%%%%%%%%%%%%%%%%%%%%%%%%%%%%%%%%%%%%%%%%%%%%%%%%%%%%%
\section{Recent discussion on ultramagnetized white dwarfs}\label{sec:8}
%%%%%%%%%%%%%%%%%%%%%%%%%%%%%%%%%%%%%%%%%%%%%%%%%%%%%%%%%%%%%%%%%%%%%%%%%%%%%%%%%%%%%%%%%%%%%%%%%%
%%%%%%%%%%%%%%%%%%%%%%%%%%%%%%%%%%%%%%%%%%%%%%%%%%%%%%%%%%%%%%%%%%%%%%%%%%%%%%%%%%%%%%%%%%%%%%%%%%

Before concluding, it is worth to mention that during the refereeing process of 
this work, several criticisms have been raised about the new mass limit for white dwarfs
presented by \citet{2013PhRvL.110g1102D}. Some of the inconsistencies of that ultramagnetized
Super-Chandrasekhar white dwarf model such as virial theorem violation, inverse $\beta$ decay
and pycnonuclear instabilities, breaking of spherical symmetry, and general relativistic effects
have been analyzed here. We refer the reader to
\citep{2013PhRvD..88h1301C,2014PhRvL.112c9001D,Sushan2014,2014PhRvD..89j3017N} for 
further details on some of the above points and for some others such as the neglected
effect of the Lorentz force (magnetic field gradient).

Very recently, \citet{Das2014} obtained new solutions for ultramagnetized, 
Super-Chandrasekhar white dwarfs, which take into account some of the above criticisms,
leading to an improvement of the treatment. They solve the general relativistic equations 
of hydrostatic equilibrium within the assumption of spherical symmetry including the magnetic
pressure gradient. The effect of the magnetic field gradient was introduced through a phenomenological 
magnetic field profile. They solved the equations for two different conditions on the parallel 
pressure: ($i$) that the spherically averaged parallel pressure be positive throughout the star, 
or ($ii$) the parallel pressure be positive throughout. The total pressure of the system was assumed
isotropic and increased by an \emph{isotropized} magnetic field contribution $(1/3)B^2/(8\pi)=B^2/(24 \pi)$.
Clearly this isotropic increase of the matter pressure could give, in principle, to systems with higher masses
with respect to an unmagnetized case, as indeed \citet{Das2014} obtained. They find that for the constraint 
($i$) the maximum mass could be (for some choice of the phenomenological parameters of the magnetic field 
profile) as high as $M_{\rm max} \approx 3.3\,M_\odot$, and for the constraint
($ii$), $M_{\rm max} \approx 2.1\,M_\odot$. The magnetic field at the center in these 
configurations is $B_{\rm center}\approx 6.8\times 10^{14}$~G. Those solutions, although
interesting, use a phenomenological magnetic field profile not coming from the self-consistent 
solution of the Maxwell equations coupled to the Einstein equations. It is not clear that the 
self-consistent solution will have a distribution of the magnetic field similar to the one 
employed and with a value showing such a high excursion from the center to the surface.
A good example for the latter is the self-consistent solution by \citet{1954Ferraro}, for
which the magnetic field at the center is only five times larger than its value at the surface.
Possibly a more self-consistent calculation has been recently performed by \citet{2014arXiv1405.2282B},
and which includes the break of the spherical symmetry and the effect of the quantum pressure anisotropy. 
They obtain white dwarf masses as large as $1.9\,M_\odot$. However, the maximum mass solution was obtained 
there for an electron Fermi energy which overcomes the limiting value for inverse $\beta$ decay analyzed in
this work and in \citep{2013PhRvD..88h1301C}.

As a positive support for their model, \citet{Das2014} recalled the recent
mathematical analysis by \citet{Federbush} who showed that there exist solutions
for magnetic self-gravitating $n=1$ polytropes for a specific ansatz of the
current $J=\beta r \rho$, where $r$ is the cylindrical coordinate, $\rho$ is
the matter density, and $\beta$ a constant. For the case of constant density
the above ansatz reduces to the one introduced by \citet{1954Ferraro}. \citet{Federbush} 
proved that there exist solutions providing the constant $\beta$ is properly bound by a
sufficiently small value. However, the solutions found by \citet{2013PhRvL.110g1102D}
and \citet{Das2014} do not conform such an ansatz of the current and therefore the analysis
of \citet{Federbush} does not apply for such a specific solution. It is noteworthy that, in 
addition, \citet{Federbush} provides a simple proof on the non-existence of magnetic stars in
the spherically symmetric case since the only possible solution has a magnetic field with a singularity at the center.

%%%%%%%%%%%%%%%%%%%%%%%%%%%%%%%%%%%%%%%%%%%%%%%%%%%%%%%%%%%%%%%%%%%%%%%%%%%%%%%%%%%%%%%%%%%%%%%%%%
%%%%%%%%%%%%%%%%%%%%%%%%%%%%%%%%%%%%%%%%%%%%%%%%%%%%%%%%%%%%%%%%%%%%%%%%%%%%%%%%%%%%%%%%%%%%%%%%%%
\section{Conclusions and discussion}\label{sec:9}
%%%%%%%%%%%%%%%%%%%%%%%%%%%%%%%%%%%%%%%%%%%%%%%%%%%%%%%%%%%%%%%%%%%%%%%%%%%%%%%%%%%%%%%%%%%%%%%%%%
%%%%%%%%%%%%%%%%%%%%%%%%%%%%%%%%%%%%%%%%%%%%%%%%%%%%%%%%%%%%%%%%%%%%%%%%%%%%%%%%%%%%%%%%%%%%%%%%%%

We have shown that the ultramagnetized, $B\gtrsim 10^{15}$ G, massive, $M\gtrsim 2 M_\odot$, white dwarfs introduced 
in \citep{2013PhRvL.110g1102D} are unlikely to exist in nature since they are subjected to several macro and micro
instabilities which would make a white dwarf either to collapse or to explode much prior to the reaching of such a 
hypothetical structure. The construction of equilibrium configurations of a magnetized compact star needs the inclusion
of several effects not accounted for in \citep{2013PhRvL.110g1102D}, and therefore the acceptance of such ultramagnetized
white dwarfs as possible astrophysical objects has to be considered with most caution. On the contrary, sub-Chandrasekhar white 
dwarfs (or slightly exceeding the Chandrasekhar limiting value e.g. by rotation) with surface magnetic fields in the observed 
range, i.e.~$B\sim 10^6$--$10^{10}$~G, can be safely described using an unmagnetized approximation for the calculation of 
the structure parameters such as mass and radius.

JGC acknowledges the support by the International Cooperation Program CAPES-ICRANet financed by
CAPES - Brazilian Federal Agency for Support and Evaluation of Graduate Education within the Ministry of Education of Brazil and MM acknowledges 
the financial support of CNPq and FAPESP (S\~{a}o Paulo state agency, thematic project $\#$2013/26258-4).

%%%%%%%%%%%%%%%%%%%%%%%%%%%%%%%%%%%%%%%%%%%%%%%%%%%%%%%%%%%%%%%%%%%%%%%%%%%%%%%%%%%%%%%%%%%%%%%%%%
%%%%%%%%%%%%%%%%%%%%%%%%%%%%%%%%%%%%%%%%%%%%%%%%%%%%%%%%%%%%%%%%%%%%%%%%%%%%%%%%%%%%%%%%%%%%%%%%%%

\bibliographystyle{apj}
\bibliography{biblio}

\begin{thebibliography}{}
\expandafter\ifx\csname natexlab\endcsname\relax\def\natexlab#1{#1}\fi

\bibitem[{{Barstow} {et~al.}(1995){Barstow}, {Jordan}, {O'Donoghue},
  {Burleigh}, {Napiwotzki}, \& {Harrop-Allin}}]{1995MNRAS.277..971B}
{Barstow}, M.~A., {Jordan}, S., {O'Donoghue}, D., {et~al.} 1995, \mnras, 277,
  971

\bibitem[{{Baym} {et~al.}(1971){Baym}, {Bethe}, \& {Pethick}}]{baym71}
{Baym}, G., {Bethe}, H.~A., \& {Pethick}, C.~J. 1971, \npa, 175, 225

\bibitem[{{Bera} \& {Bhattacharya}(2014)}]{2014arXiv1405.2282B}
{Bera}, P., \& {Bhattacharya}, D. 2014, ArXiv:1405.2282, arXiv:1405.2282

\bibitem[{{Boshkayev} {et~al.}(2013{\natexlab{a}}){Boshkayev}, {Izzo}, {Rueda
  Hernandez}, \& {Ruffini}}]{2013A&A...555A.151B}
{Boshkayev}, K., {Izzo}, L., {Rueda Hernandez}, J.~A., \& {Ruffini}, R.
  2013{\natexlab{a}}, \aap, 555, A151

\bibitem[{{Boshkayev} {et~al.}(2013{\natexlab{b}}){Boshkayev}, {Rueda},
  {Ruffini}, \& {Siutsou}}]{2013ApJ...762..117B}
{Boshkayev}, K., {Rueda}, J.~A., {Ruffini}, R., \& {Siutsou}, I.
  2013{\natexlab{b}}, \apj, 762, 117

\bibitem[{{Chaichian} {et~al.}(2000){Chaichian}, {Masood}, {P{\'e}rez
  Mart{\'i}nez}, \& {P{\'e}rez Rojas}}]{PRL2000Aurora}
{Chaichian}, M., {Masood}, S.~S., {P{\'e}rez Mart{\'i}nez}, A., \& {P{\'e}rez
  Rojas}, H. 2000, \prl, 84, 5261

\bibitem[{{Chamel} {et~al.}(2013){Chamel}, {Fantina}, \&
  {Davis}}]{2013PhRvD..88h1301C}
{Chamel}, N., {Fantina}, A.~F., \& {Davis}, P.~J. 2013, \prd, 88, 081301

\bibitem[{Chamel {et~al.}(2012)Chamel, Pavlov, Mihailov, Velchev, Stoyanov,
  Mutafchieva, Ivanovich, Pearson, \& Goriely}]{PhysRevC.86.055804}
Chamel, N., Pavlov, R.~L., Mihailov, L.~M., {et~al.} 2012, Phys. Rev. C, 86,
  055804

\bibitem[{{Chandrasekhar} \& {Fermi}(1953)}]{1953ApJ...118..116C}
{Chandrasekhar}, S., \& {Fermi}, E. 1953, \apj, 118, 116

\bibitem[{{Coelho} \& {Malheiro}(2013{\natexlab{a}})}]{Coelho2013aarxiv}
{Coelho}, J.~G., \& {Malheiro}, M. 2013{\natexlab{a}}, arXiv:1307.8158,
  arXiv:1307.8158

\bibitem[{{Coelho} \& {Malheiro}(2013{\natexlab{b}})}]{Coelho2013barxiv}
---. 2013{\natexlab{b}}, arXiv:1303.0863v2, arXiv:1303.0863v2

\bibitem[{{Coelho} \& {Malheiro}(2014)}]{Coelho2014}
---. 2014, PASJ, 66, 1

\bibitem[{{Das} \& {Mukhopadhyay}(2013)}]{2013PhRvL.110g1102D}
{Das}, U., \& {Mukhopadhyay}, B. 2013, \prl, 110, 071102

\bibitem[{{Das} \& {Mukhopadhyay}(2014)}]{Das2014}
---. 2014, arxiv:1404.7627, arXiv:1404.7627

\bibitem[{{Das} {et~al.}(2013){Das}, {Mukhopadhyay}, \&
  {Rao}}]{2013ApJ...767L..14D}
{Das}, U., {Mukhopadhyay}, B., \& {Rao}, A.~R. 2013, \apjl, 767, L14

\bibitem[{{Dong} {et~al.}(2014){Dong}, {Zuo}, {Yin}, \&
  {Gu}}]{2014PhRvL.112c9001D}
{Dong}, J.~M., {Zuo}, W., {Yin}, P., \& {Gu}, J.~Z. 2014, Physical Review
  Letters, 112, 039001

\bibitem[{{Federbush} {et~al.}(2014){Federbush}, {Luo}, \&
  {Smoller}}]{Federbush}
{Federbush}, P., {Luo}, T., \& {Smoller}, J. 2014, arXiv:1402.0265,
  arXiv:1402.0265

\bibitem[{{Ferrario} {et~al.}(2005){Ferrario}, {Wickramasinghe}, {Liebert}, \&
  {Williams}}]{2005MNRAS.361.1131F}
{Ferrario}, L., {Wickramasinghe}, D., {Liebert}, J., \& {Williams}, K.~A. 2005,
  \mnras, 361, 1131

\bibitem[{{Ferraro}(1954)}]{1954Ferraro}
{Ferraro}, V. C.~A. 1954, \apj, 119, 407

\bibitem[{{Gasques} {et~al.}(2005){Gasques}, {Afanasjev}, {Aguilera}, {Beard},
  {Chamon}, {Ring}, {Wiescher}, \& {Yakovlev}}]{2005PhRvC..72b5806G}
{Gasques}, L.~R., {Afanasjev}, A.~V., {Aguilera}, E.~F., {et~al.} 2005, \prc,
  72, 025806

\bibitem[{{Harrison} {et~al.}(1965){Harrison}, {Thorne}, {Wakano}, \&
  {Wheeler}}]{1965gtgc.book.....H}
{Harrison}, B.~K., {Thorne}, K.~S., {Wakano}, M., \& {Wheeler}, J.~A. 1965,
  {Gravitation Theory and Gravitational Collapse}

\bibitem[{{Harrison} {et~al.}(1958){Harrison}, {Wakano}, \&
  {Wheeler}}]{harrison1958}
{Harrison}, B.~K., {Wakano}, M., \& {Wheeler}, J.~A. 1958, Onzieme Conseil de
  Physique de Solvay

\bibitem[{{Hicken} {et~al.}(2007){Hicken}, {Garnavich}, \& {Prieto, et
  al.}}]{2007ApJ...669L..17H}
{Hicken}, M., {Garnavich}, P.~M., \& {Prieto, et al.}, J.~L. 2007, \apj, 669,
  L17

\bibitem[{{Howell} {et~al.}(2006){Howell}, {Sullivan}, \& {Nugent, et
  al.}}]{2006Natur.443..308H}
{Howell}, D.~A., {Sullivan}, M., \& {Nugent, et al.}, P.~E. 2006, \nat, 443,
  308

\bibitem[{{Ilkov} \& {Soker}(2012)}]{2012MNRAS.419.1695I}
{Ilkov}, M., \& {Soker}, N. 2012, \mnras, 419, 1695

\bibitem[{{Kepler} {et~al.}(2010){Kepler}, {Kleinman}, {Pelisoli}, {Pe{\c
  c}anha}, {Diaz}, {Koester}, {Castanheira}, \& {Nitta}}]{2010AIPC.1273...19K}
{Kepler}, S.~O., {Kleinman}, S.~J., {Pelisoli}, I., {et~al.} 2010, in American
  Institute of Physics Conference Series, Vol. 1273, American Institute of
  Physics Conference Series, ed. {K.~Werner \& T.~Rauch}, 19--24

\bibitem[{{Kepler} {et~al.}(2013){Kepler}, {Pelisoli}, {Jordan}, {Kleinman},
  {Koester}, {K{\"u}lebi}, {Pe{\c c}anha}, {Castanheira}, {Nitta}, {Costa},
  {Winget}, {Kanaan}, \& {Fraga}}]{2013MNRAS.429.2934K}
{Kepler}, S.~O., {Pelisoli}, I., {Jordan}, S., {et~al.} 2013, \mnras, 429, 2934

\bibitem[{{K{\"u}lebi} {et~al.}(2010{\natexlab{a}}){K{\"u}lebi}, {Jordan},
  {Euchner}, {Gaensicke}, \& {Hirsch}}]{2010yCat..35061341K}
{K{\"u}lebi}, B., {Jordan}, S., {Euchner}, F., {Gaensicke}, B.~T., \& {Hirsch},
  H. 2010{\natexlab{a}}, VizieR Online Data Catalog, 350, 61341

\bibitem[{{K{\"u}lebi} {et~al.}(2009){K{\"u}lebi}, {Jordan}, {Euchner},
  {G{\"a}nsicke}, \& {Hirsch}}]{2009A&A...506.1341K}
{K{\"u}lebi}, B., {Jordan}, S., {Euchner}, F., {G{\"a}nsicke}, B.~T., \&
  {Hirsch}, H. 2009, \aap, 506, 1341

\bibitem[{{K{\"u}lebi} {et~al.}(2010{\natexlab{b}}){K{\"u}lebi}, {Jordan},
  {Nelan}, {Bastian}, \& {Altmann}}]{2010A&A...524A..36K}
{K{\"u}lebi}, B., {Jordan}, S., {Nelan}, E., {Bastian}, U., \& {Altmann}, M.
  2010{\natexlab{b}}, \aap, 524, A36+

\bibitem[{{Liebert} {et~al.}(1983){Liebert}, {Schmidt}, {Green}, {Stockman}, \&
  {McGraw}}]{1983ApJ...264..262L}
{Liebert}, J., {Schmidt}, G.~D., {Green}, R.~F., {Stockman}, H.~S., \&
  {McGraw}, J.~T. 1983, \apj, 264, 262

\bibitem[{{Malheiro} {et~al.}(2007){Malheiro}, {Ray}, {Cuesta}, \&
  {Dey}}]{2007IJMPD..16..489M}
{Malheiro}, M., {Ray}, S., {Cuesta}, H.~J.~M., \& {Dey}, J. 2007, International
  Journal of Modern Physics D, 16, 489

\bibitem[{{Malheiro} {et~al.}(2012){Malheiro}, {Rueda}, \&
  {Ruffini}}]{2012PASJ...64...56M}
{Malheiro}, M., {Rueda}, J.~A., \& {Ruffini}, R. 2012, PASJ, 64, 56

\bibitem[{{Morini} {et~al.}(1988){Morini}, {Robba}, {Smith}, \& {van der
  Klis}}]{1988ApJ...333..777M}
{Morini}, M., {Robba}, N.~R., {Smith}, A., \& {van der Klis}, M. 1988, \apj,
  333, 777

\bibitem[{{Nityananda} \& {Konar}(2014{\natexlab{a}})}]{Sushan2014}
{Nityananda}, R., \& {Konar}, S. 2014{\natexlab{a}}, arXiv:1405.4719,
  arXiv:1405.4719

\bibitem[{{Nityananda} \& {Konar}(2014{\natexlab{b}})}]{2014PhRvD..89j3017N}
---. 2014{\natexlab{b}}, \prd, 89, 103017

\bibitem[{{Ostriker} \& {Hartwick}(1968)}]{1968ApJ...153..797O}
{Ostriker}, J.~P., \& {Hartwick}, F.~D.~A. 1968, \apj, 153, 797

\bibitem[{{Paczynski}(1990)}]{paczynski90}
{Paczynski}, B. 1990, \apjl, 365, L9

\bibitem[{{P{\'e}rez Mart{\'i}nez} {et~al.}(2003){P{\'e}rez Mart{\'i}nez},
  {P{\'e}rez Rojas}, \& {Mosquera Costa}}]{EPJ2003Aurora}
{P{\'e}rez Mart{\'i}nez}, A., {P{\'e}rez Rojas}, H., \& {Mosquera Costa}, H.~J.
  2003, The European Physical Journal C - Particles and Fields, 29, 111

\bibitem[{{P{\'e}rez Mart{\'i}nez} {et~al.}(2008){P{\'e}rez Mart{\'i}nez},
  {P{\'e}rez Rojas}, \& {Mosquera Costa}}]{IJMPD2008Aurora}
---. 2008, International Journal of Modern Physics D, 17, 2107

\bibitem[{{Provencal} {et~al.}(1998){Provencal}, {Shipman}, {H{\o}g}, \&
  {Thejll}}]{1998ApJ...494..759P}
{Provencal}, J.~L., {Shipman}, H.~L., {H{\o}g}, E., \& {Thejll}, P. 1998, \apj,
  494, 759

\bibitem[{{Rotondo} {et~al.}(2011){Rotondo}, {Rueda}, {Ruffini}, \&
  {Xue}}]{2011PhRvD..84h4007R}
{Rotondo}, M., {Rueda}, J.~A., {Ruffini}, R., \& {Xue}, S.-S. 2011, \prd, 84,
  084007

\bibitem[{{Rueda} {et~al.}(2013){Rueda}, {Boshkayev}, {Izzo}, {Ruffini},
  {Lor{\'e}n-Aguilar}, {K{\"u}lebi}, {Aznar-Sigu{\'a}n}, \&
  {Garc{\'{\i}}a-Berro}}]{2013ApJ...772L..24R}
{Rueda}, J.~A., {Boshkayev}, K., {Izzo}, L., {et~al.} 2013, \apjl, 772, L24

\bibitem[{{Scalzo} {et~al.}(2010){Scalzo}, {Aldering}, \& {Antilogus, et
  al.}}]{2010ApJ...713.1073S}
{Scalzo}, R.~A., {Aldering}, G., \& {Antilogus, et al.}, P. 2010, \apj, 713,
  1073

\bibitem[{{Schmidt} {et~al.}(1992){Schmidt}, {Bergeron}, {Liebert}, \&
  {Saffer}}]{1992ApJ...394..603S}
{Schmidt}, G.~D., {Bergeron}, P., {Liebert}, J., \& {Saffer}, R.~A. 1992, \apj,
  394, 603

\bibitem[{{Schmidt} {et~al.}(1986){Schmidt}, {West}, {Liebert}, {Green}, \&
  {Stockman}}]{1986ApJ...309..218S}
{Schmidt}, G.~D., {West}, S.~C., {Liebert}, J., {Green}, R.~F., \& {Stockman},
  H.~S. 1986, \apj, 309, 218

\bibitem[{{Shapiro} \& {Teukolsky}(1983)}]{1983bhwd.book.....S}
{Shapiro}, S.~L., \& {Teukolsky}, S.~A. 1983, {Black holes, white dwarfs, and
  neutron stars: The physics of compact objects}

\bibitem[{{Silverman} {et~al.}(2011){Silverman}, {Ganeshalingam}, \& {Li, et
  al.}}]{2011MNRAS.410..585S}
{Silverman}, J.~M., {Ganeshalingam}, M., \& {Li, et al.}, W. 2011, \mnras, 410,
  585

\bibitem[{{Strickland} {et~al.}(2012){Strickland}, {Dexheimer}, \&
  {Menezes}}]{PRD2012Debora}
{Strickland}, M., {Dexheimer}, V., \& {Menezes}, D.~P. 2012, \prd, 86, 125032

\bibitem[{{Suh} \& {Mathews}(2000)}]{2000ApJ...530..949S}
{Suh}, I.-S., \& {Mathews}, G.~J. 2000, \apj, 530, 949

\bibitem[{{Taubenberger} {et~al.}(2011){Taubenberger}, {Benetti}, \&
  {Childress, et al.}}]{2011MNRAS.412.2735T}
{Taubenberger}, S., {Benetti}, S., \& {Childress, et al.}, M. 2011, \mnras,
  412, 2735

\bibitem[{{Tuli}(2011)}]{tuli2011}
{Tuli}, J.~K. 2011, Nuclear Wallet Cards, Ed. 8

\bibitem[{{Vauclair} {et~al.}(1997){Vauclair}, {Schmidt}, {Koester}, \&
  {Allard}}]{1997A&A...325.1055V}
{Vauclair}, G., {Schmidt}, H., {Koester}, D., \& {Allard}, N. 1997, \aap, 325,
  1055

\bibitem[{{Yamanaka} {et~al.}(2009){Yamanaka}, {Kawabata}, \& {Kinugasa, et
  al.}}]{2009ApJ...707L.118Y}
{Yamanaka}, M., {Kawabata}, K.~S., \& {Kinugasa, et al.}, K. 2009, \apj, 707,
  L118

\end{thebibliography}

\end{document}